\documentclass[preprint,3p, onecolumn]{elsarticle}

\usepackage{lineno}
\modulolinenumbers[1]

\journal{Journal of \LaTeX\ Templates}

\newcommand{\tj}[6]{ \begin{pmatrix}
  #1 & #2 & #3 \\
  #4 & #5 & #6 
 \end{pmatrix}}

\newcommand{\sj}[6]{ \begin{Bmatrix}
  #1 & #2 & #3 \\
  #4 & #5 & #6 
 \end{Bmatrix}}

\newcommand{\nj}[9]{ \begin{Bmatrix}
  #1 & #2 & #3 \\
  #4 & #5 & #6 \\
  #7 & #8 & #9
 \end{Bmatrix}}
\usepackage[utf8x]{inputenc}
\usepackage{textgreek}
\usepackage{isotope}
\usepackage{subcaption}
\usepackage{mathtools} 
\usepackage{braket}
\usepackage{comment}
\usepackage[colorlinks,citecolor=blue,urlcolor=blue,linkcolor=blue]{hyperref}

\let\today\relax
\makeatletter
\def\ps@pprintTitle{%
    \let\@oddhead\@empty
    \let\@evenhead\@empty
    \def\@oddfoot{\footnotesize\itshape
         {} \hfill\today}%
    \let\@evenfoot\@oddfoot
    }
\makeatother

\begin{document}
\begin{frontmatter}

\title{Monte Carlo simulations of $\gamma$-directional correlations and their application on FIFRELIN cascades}



\author[mymainaddress]{A. Chalil\corref{mycorrespondingauthor}}
\ead{achment.chalil@cea.fr}
\author[mymainaddress]{T. Materna}
\author[mysecondaryaddress]{O. Litaize}
\author[mysecondaryaddress]{A. Chebboubi}
\author[mymainaddress]{F. Gunsing}

\cortext[mycorrespondingauthor]{Corresponding author}

\address[mymainaddress]{IRFU, CEA, Universit\'{e} Paris-Saclay, 91191 Gif-sur-Yvette, France}
\address[mysecondaryaddress]{CEA, DES, IRESNE, DER, Cadarache F-13108 Saint-Paul-Lez-Durance, France}

\begin{abstract}

Angular distribution and correlation measurements are an essential part in nuclear structure experiments, especially when spectroscopic information of a specific nucleus is unknown. In most cases, the experimental determination of the spins, parities of the studied nuclear states, as well as the possible mixing between two electric/magnetic multipoles of a transition are determined using angular correlation measurements. 
In this work, the full effect of directional $\gamma$-correlations is simulated, by using the formal theory of angular distributions. The density matrix formalism along with its multipole expansions called statistical tensors is employed, enabling to perform a full simulation of the angular correlation effects in a cascade of an arbitrary number of $\gamma$ transitions. A triple $\gamma$ angular correlation simulation is demonstrated for the first time. The present approach was coupled with the Monte Carlo code FIFRELIN, which can simulate the de-excitation of fission fragments or of excited nuclei after neutron capture. It provides a complete description of the spatial distributions of all the $\gamma$ rays in the cascade, that can be used for simulation purposes in various applications both in nuclear and particle physics. The potential for a novel approach in data analysis of angular correlation measurements is discussed thoroughly.

\end{abstract}

\begin{keyword}

angular correlations \sep
statistical tensors \sep
Monte Carlo simulations \sep
FIFRELIN
\end{keyword}

\end{frontmatter}

\section{Introduction}
\label{intro}

The correlation between successive radiations emitted by de-exciting nuclei has been theoretically predicted from the 1940's, when Hamilton~\cite{Hamilton1940_PhysRev.58.122} and Goertzel~\cite{Goertzel_1946_PhysRev.70.897} predicted that there should be an anisotropy between two successive quanta emitted from a single radiative system. The first experimental evidence was published by Brandy and Deutsch~\cite{Brandy_PhysRev.78.558}, by observing the angular correlations of successive gamma rays in six even-even nuclei. Their results have been found to be in agreement with the theoretical predictions.

In general, the anisotropy of a radiation emitted from a state with spin $J$ arises from the unequal population of the $2J+1$ magnetic $m$-substates. Biedenharn, Rose and Brink~\cite{Biedenharn_Rose_RevModPhys.25.729, Rose_Brink_Revmodphys_1967} have developed the formal theory of angular correlations, by defining the statistical tensors $\rho_q^\lambda(J)$ for a nuclear state of spin $J$, using the density matrix formalism~\cite{Fano_1957_RevModPhys.29.74}. These statistical tensors are expansions of the density matrix in the angular momentum space, and provide a very convenient means of describing the information of the unequal $m$ populations in nuclear levels. More details for the statistical tensor formalism can be found in~\cite{Hamilton1975_electromagnetic}. 

The evaluation of the statistical tensors for a cascade of two or more radiations, leads to the determination of the angular correlation function which describes the anisotropy between the emitted radiations. For a cascade of two subsequent $\gamma$-rays, the angular correlation function, i.e. the probability of detecting these two $\gamma$-rays within a relative angle $\theta_{rel}$ is determined as~\cite{Hamilton1975_electromagnetic,DEGROOT1952_1201,Kraus_1953}:
\begin{equation}\label{eq: angular correlation function}
    W(\theta_{rel})=  A_0 \left[1 +  \sum_{\substack{\lambda>0 \\ \lambda = even}} a_\lambda P_\lambda(
    \cos \theta_{rel})  \right].
\end{equation}
where $A_0$ is a normalization factor, $a_\lambda$ are coefficients that depend on the spins and multipolarities of the transitions~\cite{TAYLOR19711} and $P_\lambda$ are the ordinary Legendre polynomials. If polarization is not observed, only even values of $\lambda$ are kept in the sum. The summation is running up to twice the minimum of the spins and the multipolarities involved in the cascade ~\cite{Hamilton1975_electromagnetic,STUCHBERY200369}. The coefficients $a_\lambda$ contain the information about spins and the possible mixing ratios between the two or more types of radiation.

In the present work, a method for simulating the full effect of angular correlations is explained in detail. The particular method can be coupled to powerful codes such as FIFRELIN~\cite{Litaize_2010_PhysRevC.82.054616,LITAIZE201251,Litaize2015,REGNIER201619} and DICEBOX~\cite{BECVAR1998434}. The FIFRELIN code has been developed for the evaluation of fission data providing accurate description on the neutron and gamma properties of the fission process. FIFRELIN employs a Monte Carlo Hauser-Feshbach framework based on Bečvár’s algorithm devoted to gamma emission~\cite{BECVAR1998434} and extended to coupled neutron/gamma emission~\cite{Litaize_2010_PhysRevC.82.054616}. A sample of nuclear level schemes is generated taking into account the uncertainties from nuclear structure. FIFRELIN samples the lower energy part of the level scheme from the RIPL-3 database~\cite{CAPOTE20093107}. For the higher energy part, a combination between known levels and theoretical nuclear models (level densities, gamma strength functions, spin/parity distributions) is used, to account for the unknown part of the true level scheme of the nucleus of interest. The FIFRELIN code is also used for the simulations which are performed for the STEREO experiment~\cite{Allemandou_2018}. Recent simulations performed for the particular  experiment~\cite{Almazan2019}, which heavily rely on accurate description of the de-excitation of the Gd isotopes, have also demonstrated a better agreement of FIFRELIN with the data in comparison with GEANT4~\cite{Geant4_ALLISON2016186}. 

In addition, we discuss a novel approach for the determination of the observables that can be obtained from angular correlation or, in general, angular distribution measurements. This approach takes advantage of the Monte Carlo technique, in order to simulate the desired cascade of gammas (or other particles), according to the respective probability distributions that the formal theory provides. The extraction of a simulated angular correlation function $W_{sim}$, combined with detector simulations can be then directly compared to the experimental data. 

The important asset of this process is that triple or higher order angular correlations~\cite{Hamilton1975_electromagnetic, Singh_PhysRevC.4.1510,Krane_1988_PhysRevC.37.747,KRANE1983321,KRANE_steffen_wheeler_1973351} can be directly compared, without the derivation and use of complicated formulae. A triple $\gamma$ cascade angular correlation simulation is demonstrated in this work for the first time. Furthermore, a potential combination with the detector simulation eliminates the need for corrections due to the finite-size dimensions of the detectors~\cite{KRANE1972_solid_angle_205,  Krane1973SolidangleCF, CAMP1969192, Moura_2019_doi:10.1119/1.5099891}. Although these corrections are usually small and often not used, they can play an important role for the accurate determination of the mixing ratio. 

In the following, the method to generate the full effect of directional $\gamma$-ray angular correlations in a cascade of transitions with an arbirtrary number of $\gamma$-rays is described in detail. The well known cascade $1384\rightarrow 885 \rightarrow 658$ from the decay of \textsuperscript{110m}Ag~\cite{Singh_PhysRevC.4.1510,Krane_1988_PhysRevC.37.747} is simulated and compared to the theoretical distributions. The potential coupling of the present work with simulations performed with FIFRELIN is also discussed, and the method is applied to FIFRELIN output for the isotope \textsuperscript{156}Gd. Applications of such coupled schemes for more complicated simulations such as the STEREO experiment simulation, are also discussed.

\begin{figure*}[t]
\centering
		\includegraphics[width=0.6\textwidth]{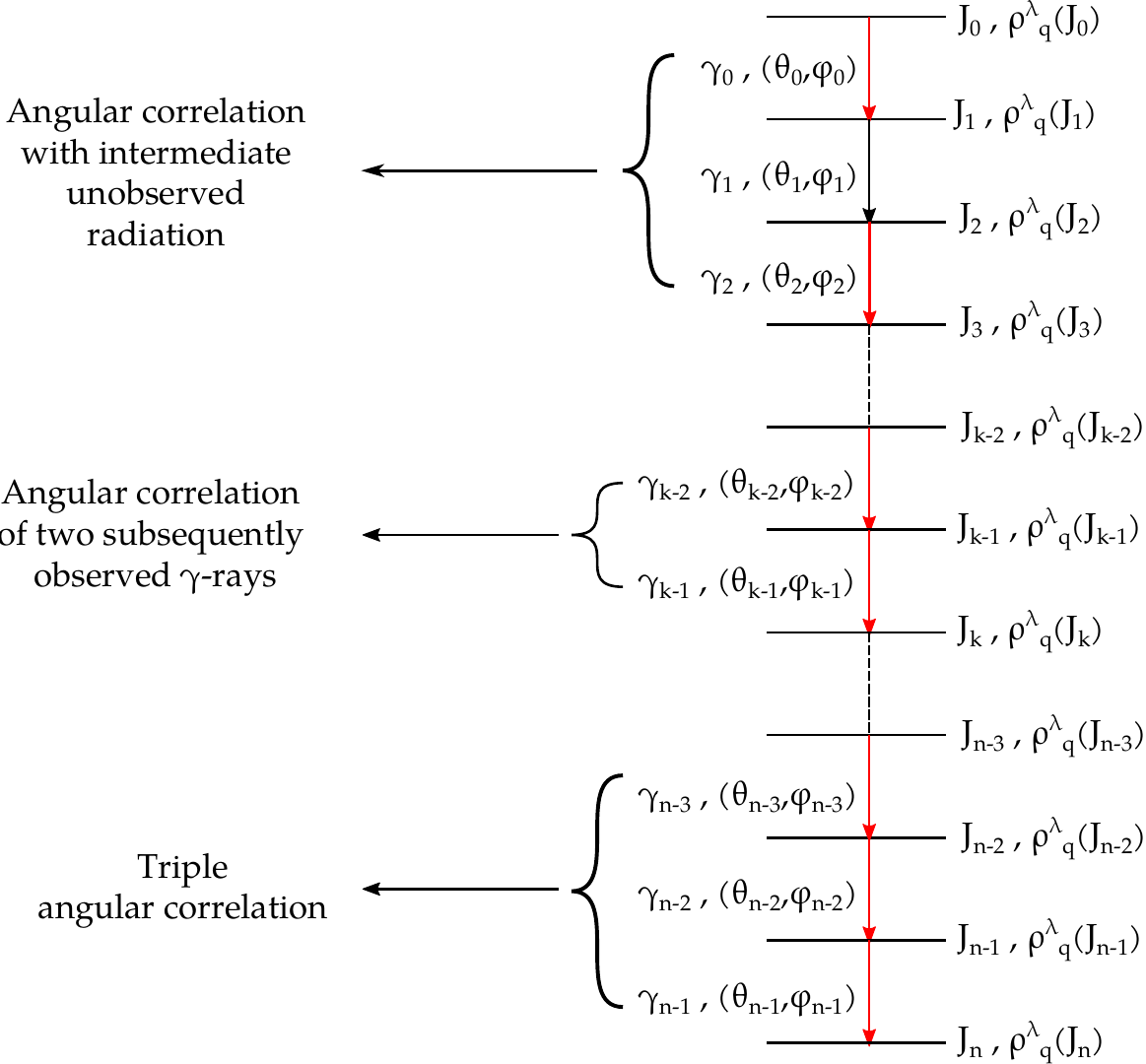}
		\centering

	\caption{[Color online] Schematic illustration of the de-excitation scheme in a cascade of $\gamma$-rays. Each state is characterized by a spin $J$ and the statistical tensor $\rho^\lambda_q(J)$, which depends on the spins of the initial and final states, as well as the multipolarities and angles of emission of the previously emitted $\gamma$-ray. Transitions marked with red illustrate that the $\gamma$ ray is observed by a detector. See text for details.}
	\label{fig:illustration}
\end{figure*}

\section{Monte carlo simulation of the angular correlations}
\label{sec:monte_carlo}

\subsection{Theoretical description}
A general method of simulating angular distributions and correlations from decaying particles with spin using the density matrix formalism~\cite{Fano_1957_RevModPhys.29.74}, has been presented in~\cite{AMSLER198321}. The event generator DECAY4~\cite{Ponkratenko2000} can simulate the effect of directional angular correlations, but the theoretical description in the article is not detailed. GEANT4~\cite{Geant4_ALLISON2016186,geant_4_sim_2017} has also the capability of simulating $\gamma$ directional correlations, by employing also the density matrix formalism, but there is no detailed documentation or step-by-step mathematical description. Furthermore, its use is restricted to radioactive sources~\cite{Turner_2020}. In addition, simulations of angular correlations have been previously made as extensions to GEANT4~{\cite{SMITH201947}}. 

For this work, the formal theory of angular distributions of radiation will be used, employing the density matrix formalism, and specifically its multipole expansions which are called statistical tensors~\cite{Rose_Brink_Revmodphys_1967,Hamilton1975_electromagnetic}. Their calculation is essential for the determination of the probability distribution functions which describe the directions of $\gamma$-rays in the cascade. For a cascade of $\gamma$-rays starting from an initial state $J_0$ and ending to a state $J_n$:
\begin{equation}\label{eq:cascade1}
    J_0 \xrightarrow[]{\gamma_0} J_1 \xrightarrow[]{\gamma_1} ...  \xrightarrow[]{\gamma_{n-1}} J_n
\end{equation}
a set of statistical tensors can be calculated, depending on the orientation of the initial state $J_0$. The set of transitions $\gamma_0$, $\gamma_1...\gamma_{n-1}$ can be either all observed or partly unobserved by a detector, as illustrated in Fig.~\ref{fig:illustration}. If the initial state is randomly oriented, as in the case of $\beta$-decay where the $\beta$-particle is not observed or an ($n,\gamma$) reaction with thermal neutrons~\cite{Snelling_1983}, then the statistical tensor takes the simple form~\cite{Hamilton1975_electromagnetic,STUCHBERY200369}:
\begin{equation}\label{eq: initial statistical tensor}
    \rho^{\lambda}_q (J_0)= \delta_{\lambda 0} \delta_{q 0}
\end{equation}
where $\delta_{ij}$ is the Kronecker delta, $\lambda$ is the rank of the statistical tensor and q takes integer values between $-\lambda$ and $\lambda$. If the initial state $J_0$ is oriented, i.e. after a nuclear reaction, the initial statistical tensor must be computed in the framework of the particular reaction or be measured experimentally. 

By knowing the initial-state statistical tensor, all the statistical tensors can then be calculated from the recursive master equation~\cite{Hamilton1975_electromagnetic,STUCHBERY200369}:
\begin{align}\label{eq: statistical tensor next simplified}
\rho^{\lambda_f}_{q_f} (J_f) &= \sum_{\lambda_i, \lambda, q, q_i} (-1)^{\lambda_i+q_i} \sqrt{2 \lambda +1}\; \rho^{\lambda_i}_{q_i} (J_i) 
 \tj{\lambda_f}{\lambda}{\lambda_i}{-q_f}{q}{q_i}  A^{\lambda_i \lambda_f}_\lambda(J_f,J_i,L,L') \mathcal{D}^{\lambda *}_{ q 0}(\phi_i,\theta_i,0)
\end{align}
where $\tj{\lambda_f}{\lambda}{\lambda_i}{-q_f}{q}{q_i}$ is the Wigner $3j$ symbol~\cite{abramovitz_handbook_1964,Wigner1993}, $A^{\lambda_i \lambda_f}_\lambda$ is the generalized angular distribution coefficient and $\mathcal{D}^{\lambda *}_{ q 0}(\phi_i,\theta_i,0)$ is a Wigner D-matrix~\cite{Wigner1993}. For the summation indices, if polarization is not observed, $\lambda_{i,f} = 0, 2, 4, . . ., 2I_{i,f}$ for integer values of $I_{i,f}$, while $\lambda_{i,f} = 0, 2, 4,. . . , 2I_{i,f} − 1$ for half-integer values of $I_{i,f}$. Also, $\lambda= 0, 2, 4 .., 2L_{max}$, where $L_{max}$ is the maximum of two lowest multipolarities $L, L'$ of the radiation. The index $q$ takes integer values within the interval $[-\lambda,+\lambda]$. The same applies to $q_i$ and $q_f$, with respect to $\lambda_i$ and $\lambda_f$. The generalized angular distribution coefficients are given by the relation:
\begin{align}\label{eq: generalized ang_distro coeff}
    A^{\lambda_i \lambda_f}_\lambda(J_f,J_i,L,L') &= \frac{1}{1+\delta^2} [ F^{\lambda_i \lambda_f}_\lambda(L,L,J_f, J_i) + 2\delta F^{\lambda_i \lambda_f}_\lambda(L,L'J_f, J_i) + \delta^2 F^{\lambda_i \lambda_f}_\lambda(L',L',J_f, J_i) ]
\end{align}
where $\delta$ is the multipolarity mixing ratio of the two lowest multipolarities $L,L'$ of the $\gamma$-transition~\cite{KRANE1975_mixing_ratio_383}. The generalized $F$-coefficients, $F^{\lambda_i \lambda_f}_\lambda$, are functions of Wigner $3j$ and $9j$ symbols~\cite{abramovitz_handbook_1964, Wigner1993} and have been extensively tabulated in literature \cite{Hamilton1975_electromagnetic,KRANE_steffen_wheeler_1973351}. They can be calculated using the relationship:
\begin{align}\label{eq: generalized F coeff}
    F^{\lambda_i \lambda_f}_\lambda(L,L',J_i,J_f) &= (-1)^{L'+\lambda_i+\lambda_f+\lambda}  \sqrt{(2J_i+1)(2J_f+1)(2L+1)(2L'+1)} \nonumber \\
     &\times \sqrt{(2\lambda_i+1)(2\lambda_f+1)(2\lambda+1)}  \tj{L}{L'}{\lambda}{1}{-1}{0}  \nj{J_f}{L}{J_i}{J_f}{L'}{J_i}{\lambda_f}{\lambda}{\lambda_i}
\end{align}
The Wigner D-Matrix is related to the spherical harmonics and the associated Legendre polynomials as~\cite{STUCHBERY200369,KRANE_steffen_wheeler_1973351}:
\begin{equation}
    \mathcal{D}^{\lambda *}_{ q 0}(\phi,\theta,0) = (-1)^q \sqrt{ \frac{4 \pi}{2 \lambda+1}}\; Y^{\lambda}_{-q} (\theta, \phi) = (-1)^{(q+|q|)/2} \sqrt{ \frac{(\lambda-|q|)!}{(\lambda+|q|)!}}\; P_\lambda^{|q|}(\cos \theta)\; e^{-iq \phi}.
\end{equation}

The description of the populations of the $m-$substates with the use of statistical tensors provides the advantage that all the populations of the subsequent states can be calculated using the master equation (Eq.~\ref{eq: statistical tensor next simplified}) and the direction $(\theta_i,\phi_i)$ of the emitted $\gamma$-ray. The directions are defined with respect to the frame of reference of the initial statistical tensor $ \rho^{\lambda}_q (J_0)$.

\subsection{Generation of the events}

After evaluating the statistical tensor for a state in the cascade, the two-dimensional angular distribution function can be evaluated by summing over the quantum numbers of the next state~\cite{Hamilton1975_electromagnetic, STUCHBERY200369}, which at this part of the simulation process is considered unobserved (i.e. the $\gamma$-ray coming from the final state has not being emitted yet in the simulation).
For the case of directional correlations, the angular distribution function is given by the relation:
\begin{align}\label{eq: distribution function}
       W(\theta_i,\phi_i) &= \sum_{\lambda_f, q_f} (-1)^{\lambda+q_f} \sqrt{2 \lambda_f +1} \rho^{\lambda_f}_{q_f} (J_f)  A_{\lambda_f}(L,L',J_i,J_f,\delta) \mathcal{D}^{\lambda_f *}_{ q_f 0}(\phi_i,\theta_i,0)
    \end{align}
where the coefficient $A_{\lambda_f}(L,L',J_i,J_f,\delta)$ is called \textit{angular distribution coefficient} and is given by the relationship~\cite{Hamilton1975_electromagnetic}:
\begin{align}\label{eq: angular distribution coefficients}
    A_\lambda(L,L',J_i,J_f,\delta) &= \dfrac{1}{1+\delta^2} \left[ F_\lambda(L,L,J_i,J_f) +  2 \delta F_\lambda(L,L',J_i,J_f)+ \delta^2 F_\lambda(L',L',J_i,J_f) \right] 
\end{align}
In the above equations, the $F$-coefficients are given by the formula~\cite{Hamilton_1985,Ferentz_F_coeff}:
\begin{equation}\label{eq: F-coefficient}
F_\lambda(L,L',J_i,J_f) = (-1)^{J_i+J_f-1}  \sqrt{ (2\lambda+1)(2L+1)(2 L'+1)(2 J_i+1) } \tj{L}{L'}{\lambda}{1}{-1}{0}
  \sj{L}{L'}{\lambda}{J_i}{J_i}{J_f};
\end{equation}
where $\sj{L}{L'}{\lambda}{J_i}{J_i}{J_f}$ is the Wigner $6j$ symbol \cite{abramovitz_handbook_1964,Wigner1993}. Thus, the simulation of the angular correlations or distributions can be performed by generating a set of directions from each of the calculated distribution functions, while moving down the cascade. The statistical tensors have to be calculated each time, after the generation of each event since they depend on the angles of emission of $\gamma$-ray that feeds the current state, as shown in Eq.~\ref{eq: statistical tensor next simplified}. The result of Eq.~\ref{eq: distribution function} should be always real for directional correlations, as it constitutes an observable quantity of the decaying nucleus.

\section{Results}
\label{sec:application}


\subsection{Double \texorpdfstring{$\gamma$}--cascades}
\begin{figure*}[t!]
\centering
	\begin{subfigure}[c]{0.50\textwidth}
		\includegraphics[width=\textwidth]{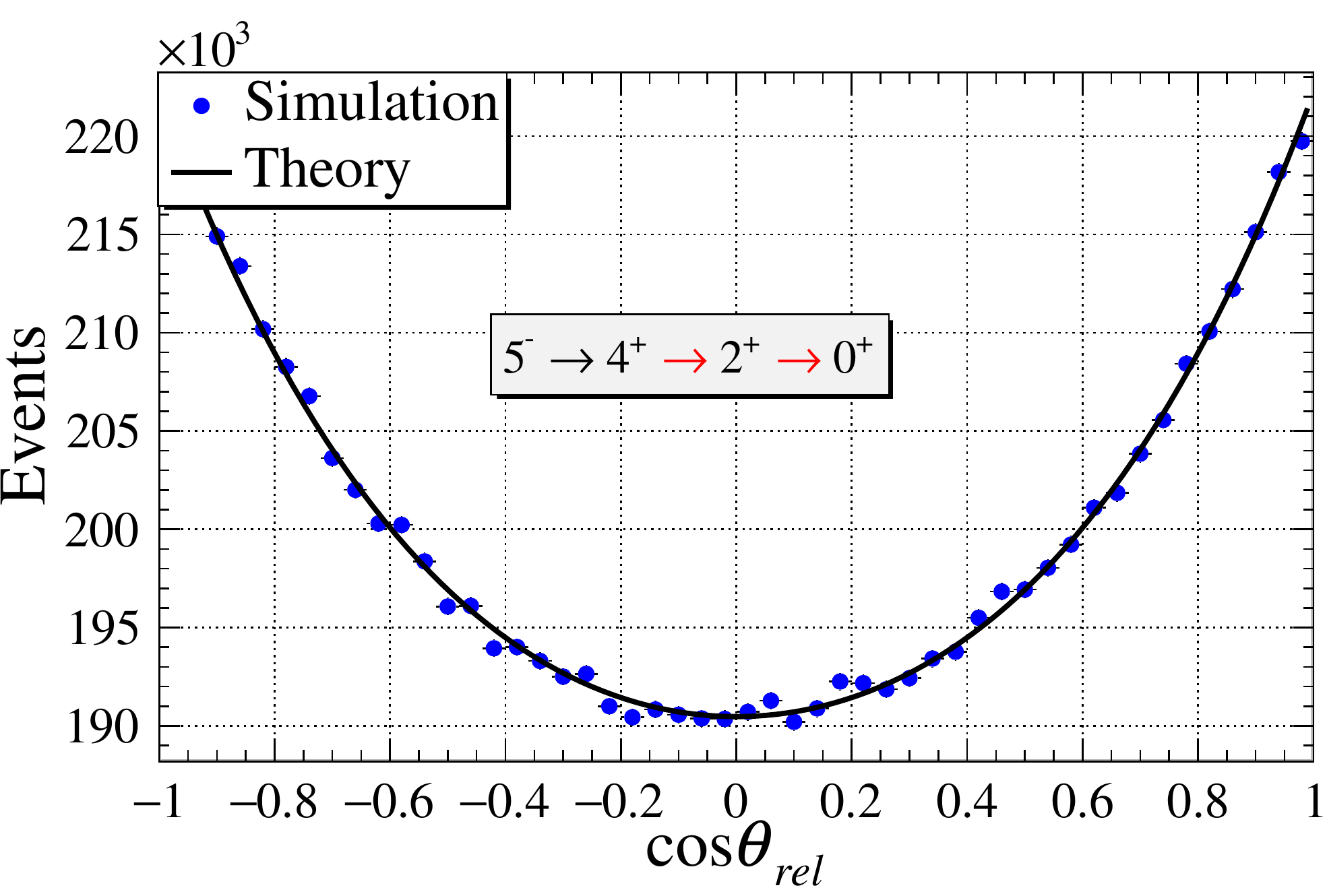}
		\centering
		\caption{}
		\label{fig: 420}
	\end{subfigure}%
	\begin{subfigure}[c]{0.50\textwidth}
		\includegraphics[width=\textwidth]{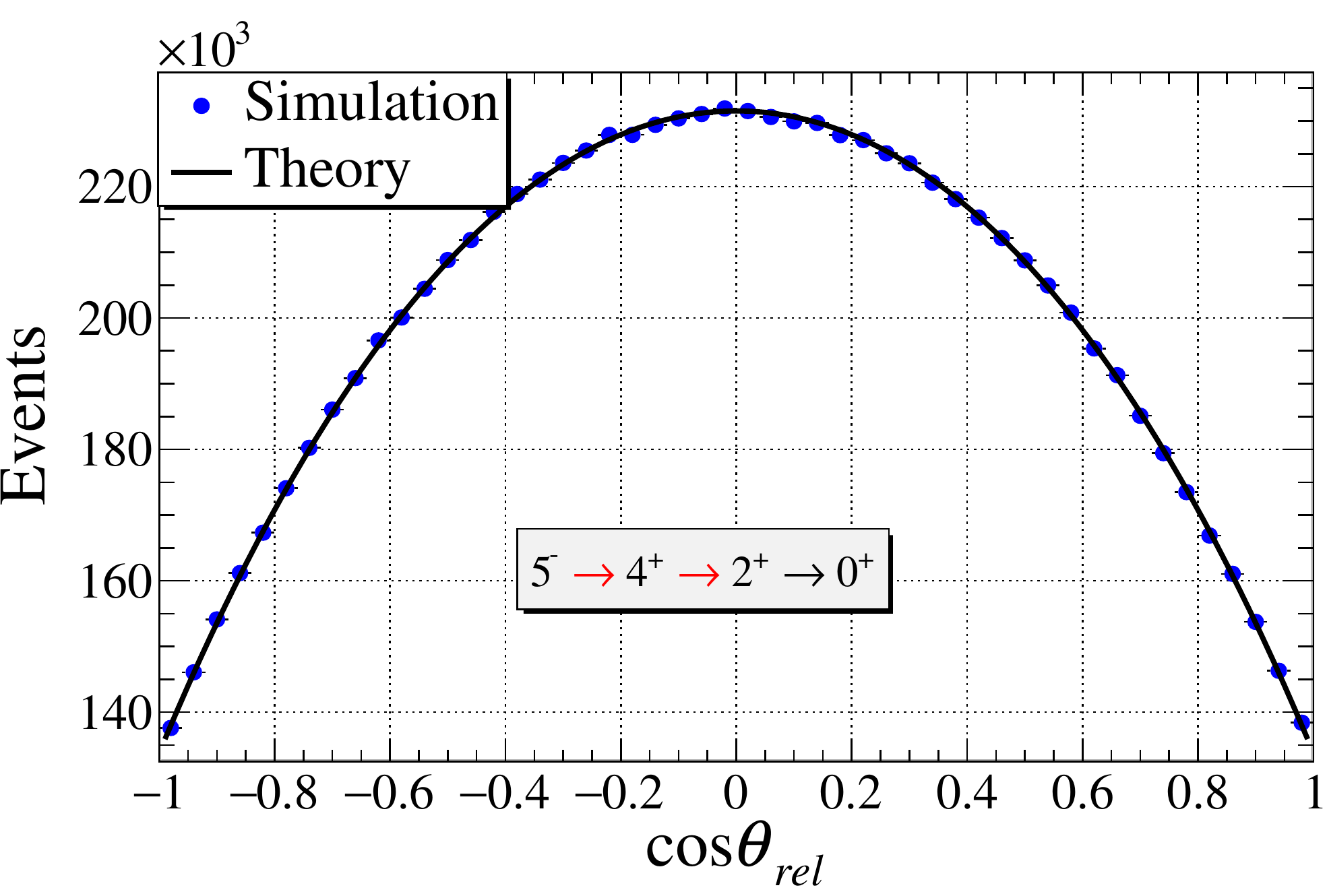}
		\centering
		\caption{}
		\label{fig: 542}
	\end{subfigure}
	\begin{subfigure}[c]{0.50\textwidth}
		\includegraphics[width=\textwidth]{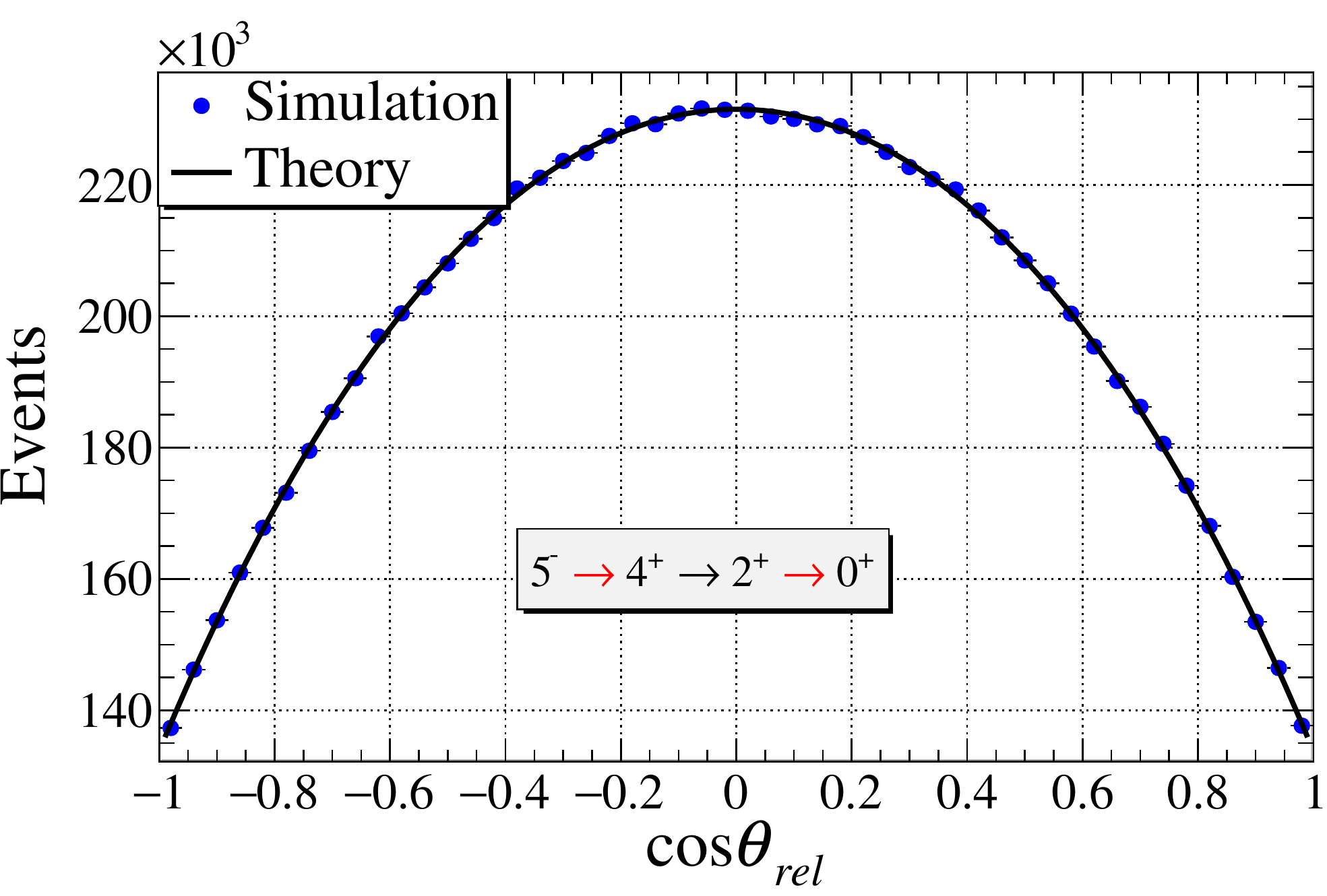}
		\centering
		\caption{}
		\label{fig: 540_un}
	\end{subfigure}%
	\begin{subfigure}[c]{0.50\textwidth}
		\includegraphics[width=\textwidth]{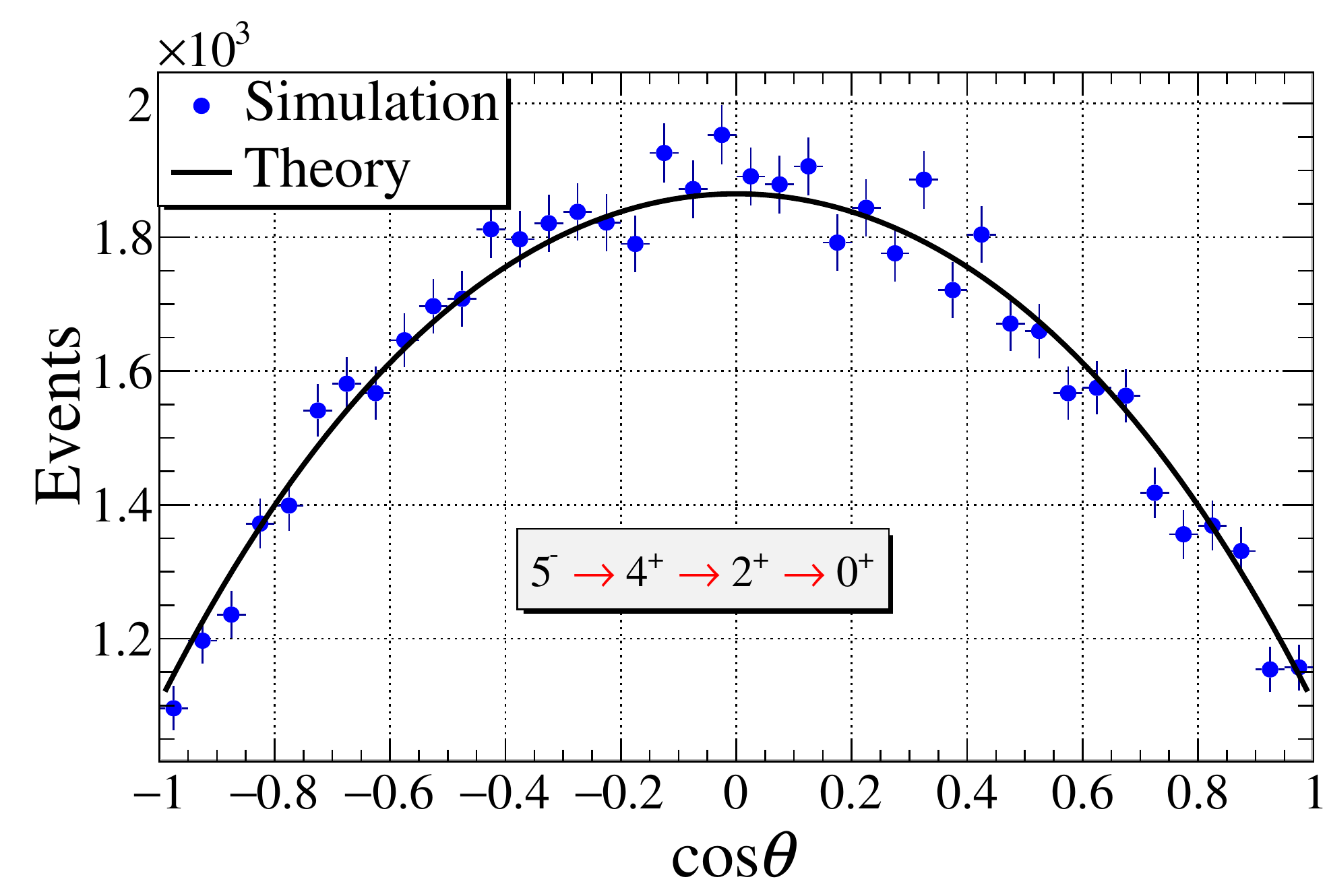}
		\centering
		\caption{}
		\label{fig: 5420triple}
	\end{subfigure}
\caption{[Color online] Monte-Carlo simulation of the $\gamma$ directional correlations for the cascade $5^- \rightarrow 4^+ \rightarrow 2^+ \rightarrow 0^+$. In (a), the distribution of relative angles for the cascade $4^+ \rightarrow 2^+ \rightarrow 0$ is shown. In (b), the distribution of relative angles for the transitions $5^- \xrightarrow[]{\delta = -0.42} 4^+ \rightarrow 2^+$ is shown. In (c), the angular correlation between the first and third transition of the same cascade is shown. The intermediate transition is treated as unobserved. In (d), the triple $\gamma$ correlation for the same cascade for the $N1$ geometry is shown (see text for details). All theoretical curves are also shown. The red arrows indicate which transitions are considered as observed.}
	\label{fig: cascade}
\end{figure*}
A simulation of a cascade of three $\gamma$-rays was performed using the present framework. The $5^- \rightarrow 4^+ \rightarrow 2^+ \rightarrow 0^+$ cascade in $^{110}$Cd, with a mixing ratio of $\delta_0 = -0.42$ for the first transition ($5^- \rightarrow 4^+$), has been extensively studied in~\cite{Krane_1988_PhysRevC.37.747}. The results of the simulation are shown in Fig~\ref{fig: cascade}. 

The simulation generates a pair of coordinates $(\theta_i, \phi_i)$ for each transition $i$. In order to check that the simulation gives the results expected from theory, a direct comparison can be made, by distributing the relative angles of two transitions. The scalar product between the two directions of the two consecutive $\gamma$ rays can be calculated and distributed to a histogram as shown in Fig.~\ref{fig: cascade}. In Fig~\ref{fig: 420}, the distribution of relative angles of the last two $\gamma$ rays is shown. This is a common  $4^+ \rightarrow 2^+ \rightarrow 0^+$ cascade, where the two $\gamma$-rays are of pure $E2$ character. The distribution of relative angles is given by the angular distribution function (Eq.~\ref{eq: angular correlation function}), where the coefficients $a_\lambda$ can be calculated using the relationship:
\begin{equation}\label{eq: a2a4coefficients}
       a_\lambda = B_\lambda(L,L',J_f,J_i,\delta_0) A_\lambda(L,L',J_i,J_f,\delta_1) 
    \end{equation}
where the functions $B_\lambda(L,L',J_f,J_i)$ are called \textit{orientation parameters} and can be calculated using the formula~\cite{Hamilton1975_electromagnetic, Blin-Stoyle1957}:
\begin{align}\label{eq: orientation parameters}
    B_\lambda(L,L',J_f,J_i,\delta_0) &= \dfrac{1}{1+\delta_0^2} \left[ F_\lambda(L,L,J_f,J_i) + (-1)^{L+L'} 2 \delta_0 F_\lambda(L,L',J_f,J_i) + \delta_0^2 F_\lambda(L',L',J_f,J_i)\right] 
\end{align}
while the coefficients $A_\lambda(L,L',J_i,J_f,\delta_1)$ are the angular distributions coefficients and can be calculated from Eq.~\ref{eq: angular distribution coefficients}.

Note that the right-hand side of the expressions in Eq. \ref{eq: angular distribution coefficients} and Eq. \ref{eq: orientation parameters} are not entirely identical, but they differ by a phase factor $(-1)^{L+L'}$. 
For the specific case of the $4^+ \rightarrow 2^+ \rightarrow 0^+$ cascade, where the transitions are of pure $E2$ character and thus $\delta_0,\delta_1 = 0$, the results for $a_2$ yield:
\begin{align}\label{eq: a2_420}
    a_2 &= B_2(2,3,2,4,0) A_2(2,2,2,0,0) = (-0.171) \times (-0.598) = 0.102
\end{align}
while for $a_4$:
\begin{align}\label{eq: a4_420}
    a_4 &= B_4(2,3,2,4,0) A_4(2,2,2,0,0) = 0.009
\end{align}
The above calculations agree with the tabulated results~\cite{TAYLOR19711} as well as with the very reliable on-line angular correlation calculator developed by the Griffin collaboration~\cite{ang_corr_calc}. The theoretical curve given by Eq.~\ref{eq: angular correlation function} is shown in Fig.~\ref{fig: 420}, along with the simulation results. In order to check the agreement, a $\chi^2$-test has been performed between the theoretical and simulated values. The value of the reduced $\chi^2$ yields $\chi^2/NDF = 1.30$. The value is close to 1, which indicates a very good agreement between the theory and the simulation. The same is shown for the upper part of the cascade ($5^- \xrightarrow{-0.42} 4^+ \rightarrow 2^+ $) in Fig.~\ref{fig: 542}. The simulation is in complete agreement with the theoretical values. The value of the reduced $\chi^2/NDF = 0.86$, indicating again a very good agreement. 




\subsection{Simulation of a double \texorpdfstring{$\gamma$}--cascade with intermediate unobserved radiations}

The present simulation is able to reproduce the angular correlation between two $\gamma$-radiations, with one or more intermediate $\gamma$-rays that are not observed by a detector. This can be checked by histogramming the scalar product of two $\gamma$-rays of interest in cascade of more than two transitions. The same cascade as before is tested, but now the distribution of relative angle between the first and the third $\gamma$-rays is studied. 

As in the case of two consecutive $\gamma$-rays, the angular correlation function~(Eq.~\ref{eq: angular correlation function}) is used to describe the angular correlation between $\gamma$-rays with intermediate unobserved radiations. The difference here is that the evaluation of the theoretical coefficients $a_\lambda$ requires the inclusion of a $U(L,L',J_i,J_f,\delta)$ factor called \textit{de-orientation coefficient}~\cite{Hamilton1975_electromagnetic}, to account for any intermediate unobserved transitions. Thus, for the specific case of the transitions $5^- \rightarrow 4^+$ and $  2^+ \rightarrow 0^+ $, where the intermediate transition ($4^+ \rightarrow 2^+$) is treated as unobserved, the coefficients $a_2, a_4$ are computed as follows:
\begin{align}\label{eq: a2a4coefficients_deorientation}
       a_\lambda &= B_\lambda(1,2,5,5,-0.42) U_\lambda(2,3,4,2,0) A_\lambda(2,2,2,0,0) 
    \end{align}
which yield $a_2 = -0.322$ and $a_4 = -0.009$. The theoretical angular correlation function, along with the results of the simulation is plotted in Fig~\ref{fig: 540_un}. A $\chi^2$-test between the theoretical curve and simulated value, yields $\chi^2/NDF = 1.25$, demonstrating the very good agreement between simulation and theory. It is important to note, that this approach works for any number of intermediate unobserved transitions. The number of de-orientation coefficients used in determining the angular correlation coefficients is the same as the number of unobserved intermediate transitions.

\subsection{Triple--\texorpdfstring{$\gamma$}~~correlations}

The case of three observed $\gamma$-rays in a cascade is also reproduced within the present framework. The determination of the parameters that describe a $\gamma \gamma \gamma$ correlation are extensively described in~\cite{KRANE_steffen_wheeler_1973351}. In general, the three $\gamma$ rays can be detected in specific geometries, such as the N1 or N2 geometries described in~\cite{KRANE_steffen_wheeler_1973351} and in~\cite{Krane_1988_PhysRevC.37.747}. 

To demonstrate the ability of simulating triple cascades, the same cascade will be now simulated, but only the events that satisfy the N1 geometry, as described in~\cite{Krane_1988_PhysRevC.37.747} will be taken into consideration. More specifically, if we consider a cartesian coordinate frame $xyz$, the N1 geometry requires the first $\gamma$ to be emitted along the $z$-axis and the second $\gamma$ in the $x$ axis. Then, the distribution of the polar angle of the third $\gamma$-ray along the $y$ axis can be determined from the distribution function~\cite{KRANE_steffen_wheeler_1973351,Hamilton1975_electromagnetic}:
\begin{equation}\label{eq: triple angular correlation function}
    W(\theta)= \Gamma_0 \left[1 +  \sum_{\substack{\lambda>0 \\ \lambda = even}} \Gamma_\lambda P_\lambda(
    \cos \theta)  \right]..
\end{equation}
where $\Gamma_0$ is a normalization factor, and $\Gamma_\lambda$ are coefficients that depend on spins and mixing ratios involved in the triple cascade. It is important to note, that the coefficients $\Gamma_\lambda$ are defined differently from the coefficients $a_\lambda$ for double cascades, which appear in Eq.~\ref{eq: angular correlation function}.

The coefficients $\Gamma_\lambda$ for the previously simulated triple cascade $5^- \xrightarrow{-0.42} 4^+ \rightarrow 2^+ \rightarrow 0+$ have been calculated theoretically in~\cite{Krane_1988_PhysRevC.37.747}, yielding $\Gamma_2 = -0.307$ and $\Gamma_4 = -0.013$. 

The simulated triple correlation along with the theoretical result is shown in Fig.~\ref{fig: 5420triple}. As illustrated, there is a very good agreement between the theory and the simulation. It has to be noted that the comparison involved a simulation with a large number of events emitted in all space ($\sim 800$ million). The comparison was made by selecting the events which are close to the particular geometry . In detail, the distribution of polar angle $\theta_2$ was accepted as a simulation event if and only if the directions of the three $\gamma$ rays $\gamma_0,\gamma_1,\gamma_2$ fall into these limits: $\theta_0 < 0.1$, $1.47 < \theta_1< 1.67$, $\phi_1<0.1$ and $\phi_1>6.18$, and $1.47< \phi_2<1.67$, where the angle intervals are in radians. The value of $\chi^2/NDF$ yields 1.09. Furthermore, the theoretical calculations performed in~\cite{Krane_1988_PhysRevC.37.747} involved also the solid angle correction factors, but their effect is expected to be small, concerning the distances mentioned in the same work.

Correlations of higher order than the triple $\gamma$-cascade have not been explored in literature. However, the results until now demonstrate the ability of simulating the complete effect of the correlation of $\gamma$-rays, including correlations that are still unexplored.

\subsection{Coupling with FIFRELIN: The case of \texorpdfstring{\textsuperscript{156}}~Gd}

\begin{figure*}[t]
\centering
	\begin{subfigure}[c]{0.5\textwidth}
		\includegraphics[width=\textwidth]{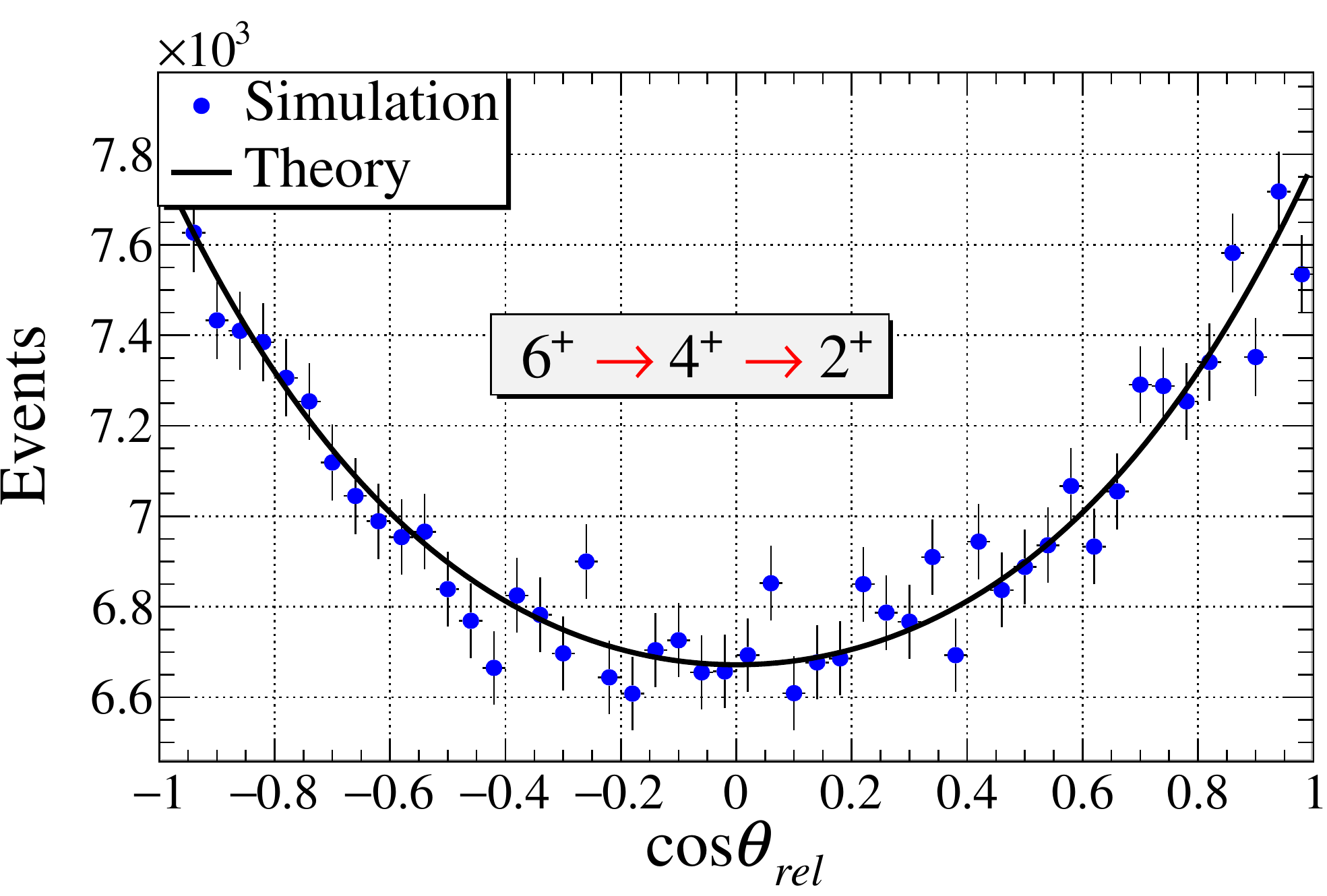}
		\centering
		\caption{}
		\label{fig: 642Gd}
	\end{subfigure}%
	\begin{subfigure}[c]{0.5\textwidth}
		\includegraphics[width=\textwidth]{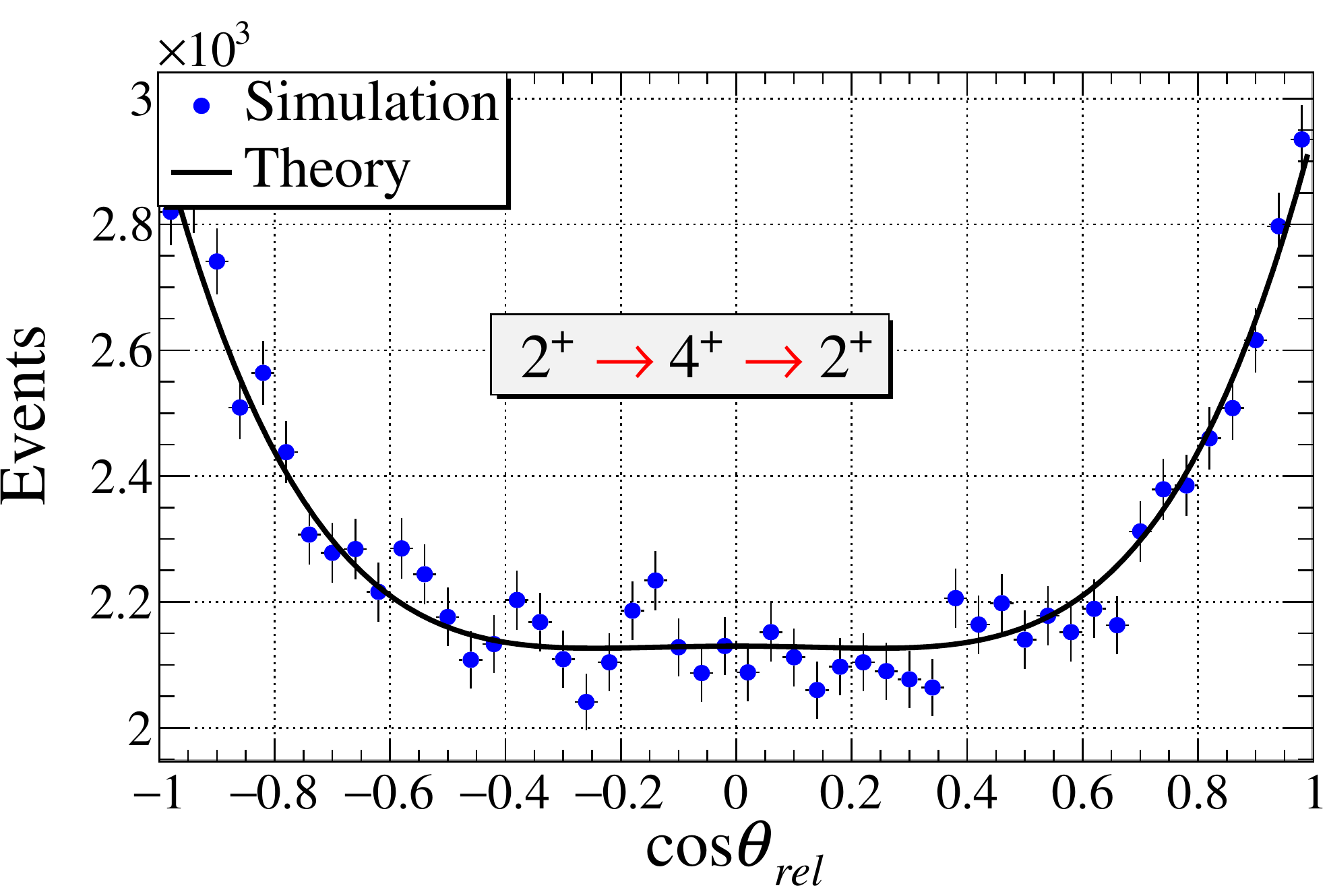}
		\centering
		\caption{}
		\label{fig: 420Gd}
	\end{subfigure}
	\begin{subfigure}[c]{0.5\textwidth}
		\includegraphics[width=\textwidth]{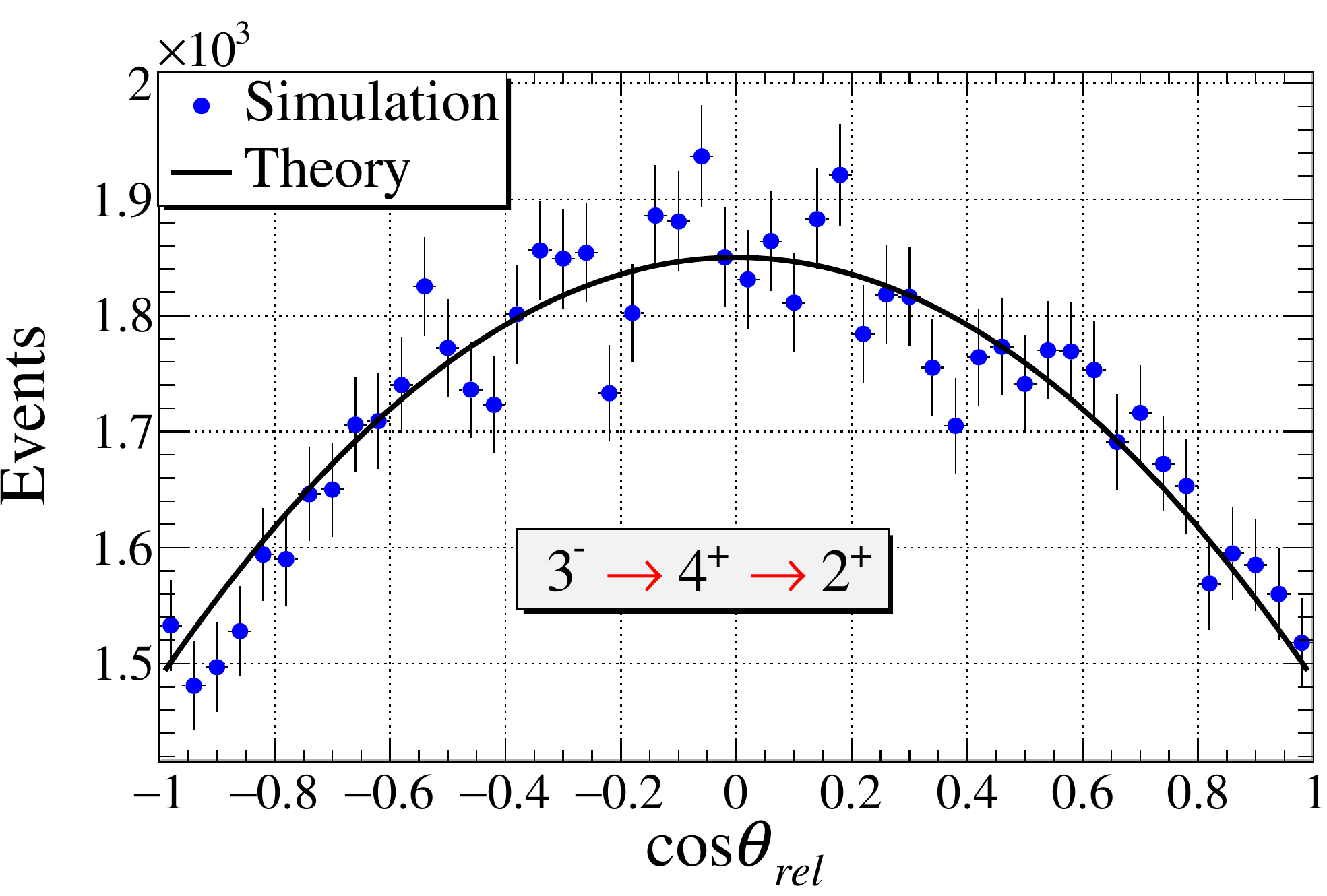}
		\centering
		\caption{}
		\label{fig: 020Gd}
	\end{subfigure}%
	\begin{subfigure}[c]{0.5\textwidth}
		\includegraphics[width=\textwidth]{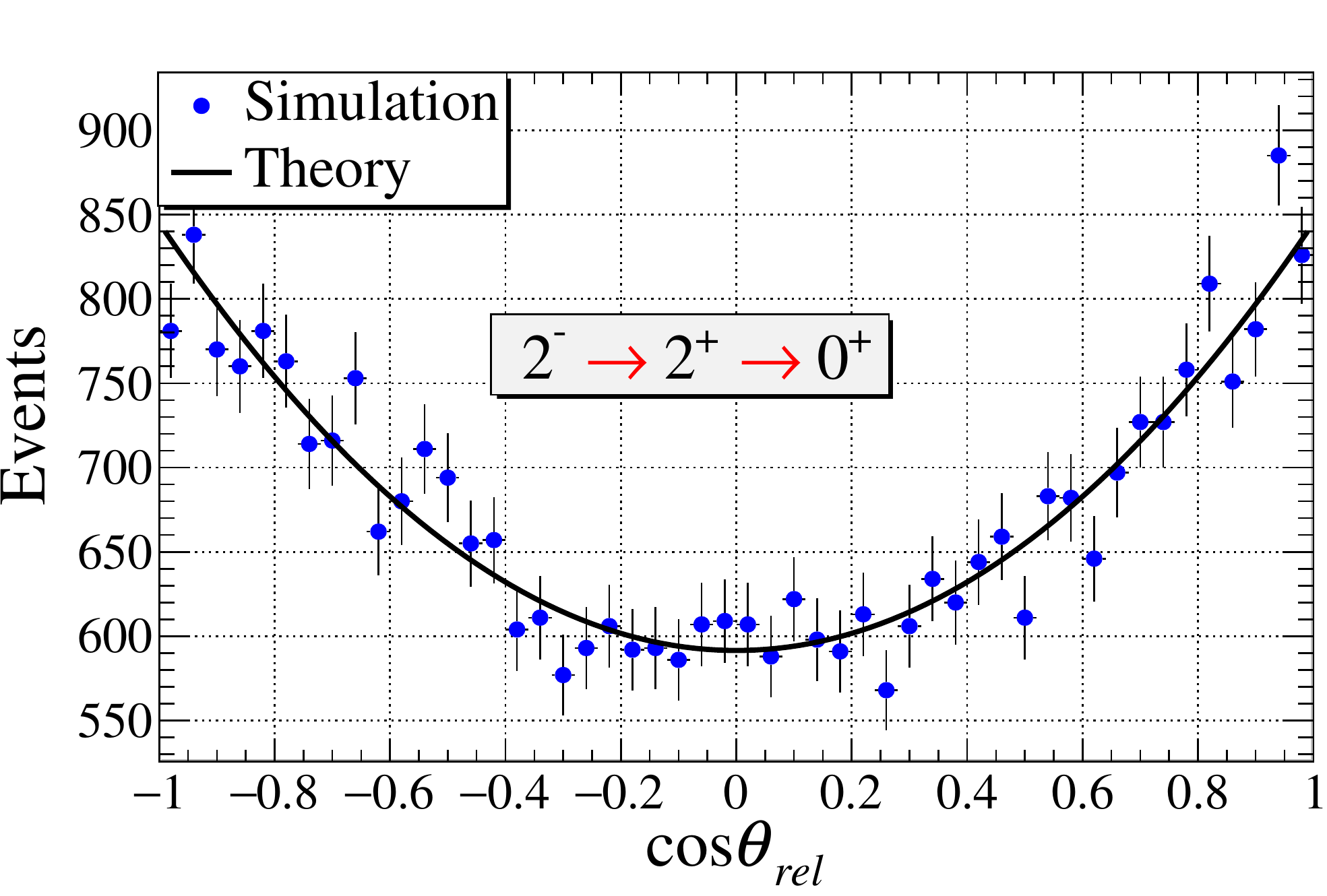}
		\centering
		\caption{}
		\label{fig: 220Gd}
	\end{subfigure}
\caption{[Color Online] (a) The angular correlation of the ground-state band cascade $6^+_1 \rightarrow 4^+_1 \rightarrow 2^+_1$ of \textsuperscript{156}Gd, along with its theoretical distribution. (b) The same for the cascade $2^+_4 \rightarrow 4^+_1 \rightarrow 2^+_1$, (c) the cascade $3^-_1 \rightarrow 4^+_1 \rightarrow 2^+_1$ and (d) the cascade $2^-_2 \rightarrow 2^+_3 \rightarrow 0^+_{g.s.}$.}
\label{fig: 156Gd cascades}
\end{figure*}

The present method has been coupled with the output of FIFRELIN code for the case of \textsuperscript{156}Gd. The accurate description of the de-excitation of the specific isotope of Gd is important for the accurate simulation of the STEREO experiment. For this case, 10 million cascades of the particular isotope were generated by FIFRELIN. The present framework was then applied on the output and the results for four cases of double correlations are presented in Fig.~\ref{fig: 156Gd cascades}.

The cascade $6^+ \rightarrow 4^+ \rightarrow 2^+$ of the rotational ground-state band of \textsuperscript{156}Gd is shown in the upper part of the figure along with the respective theoretical calculation. The particular cascade consists of two stretched $E2$ transitions and a typical theoretical calculation yields for the correlation coefficients $a_2 = 0.102$ and $a_4 = 0.009$. In the upper right part, the cascade $2^+_4 \rightarrow 4^+_1 \rightarrow 2^+_1$ is shown, which consists of two pure $E2$ transitions. The theoretical calculation for this spin sequence yields $a_2 = 0.200$ and $a_4 = 0.093$. In the bottom left part, the cascade $3^- \rightarrow 4^+ \rightarrow 2^+$ is shown, consisting of an $E1$ and an $E2$ transition. In this case the theoretical results are $a_2 = -0.14$ and $a_4 = 0$. Lastly, in the bottom right part of the figure, the $2^-_2 \rightarrow 2^+_3 \rightarrow 0^+_{g.s.}$ cascade is illustrated, where theoretical calculations result
in $a_2=0.25$ and $a_4=0$. In all cases, the simulation results are in full agreement with the theoretical calculations.

\section{Discussion}
\label{sec:discussion}


\subsection{Discussion of the method}

Within the framework of the present work, the full effect of $\gamma$-ray directional correlations is reproduced for a nuclear cascade with an arbitrary number of $\gamma$-transitions. The results presented for this study show the power of the formal theory of angular correlations to reproduce the complete effect, by producing a set of directions $(\theta_i, \phi_i)$ for each $i$-th $\gamma$-ray. 

The present simulation is able to describe all the double $\gamma$-correlations in the cascade. Double $\gamma$-correlations are the most widely used, as their dependance on spins and mixing ratios is relatively simple, enabling them to be used for the experimental determination of spins and mixing ratios on nuclei with limited spectroscopic information. The effect is demonstrated in Fig.~\ref{fig: cascade} and can be generalized for any cascade. Of course, the simulation works even for odd-mass nuclei, whose spins are half-integer.

Occasionaly, it is more convinient to measure a double $\gamma$-angular correlation with one or more intermediate unobserved radiations. Mathematically, this procedure includes an integration over positions of the intermediate $\gamma$, as the position is treated as an unobserved property of the system. The main reason for the study of such cascades aims to avoid problems in the data analysis. For example, the coincident counts of two consecutive transitions may overlap with background, making impossible the determination of an angular correlation between two consecutive $\gamma$-rays. In Fig.~\ref{fig: 540_un}, the effect is nicely demonstrated and in complete agreement with the theoretical distributions. The present method works for an arbitrary number of intermediate transitions.

Higher order angular correlations can be also reproduced. A triple $\gamma$ angular correlation is shown in Fig.~\ref{fig: 5420triple}, with the first and second transition observed in fixed direction, while the direction of the third transition is variable, according to the N1 geometry, as defined in~\cite{KRANE_steffen_wheeler_1973351}. The potential of triple $\gamma$-ray directional correlations is explored in~\cite{KRANE1983321}, where has is been demonstrated that the coincidental observation of three $\gamma$-rays instead of two can lead to a more accurate determination of multipolarity mixing ratios. The main reason for this is that the sensitivity of the angular correlation coefficients $\Gamma_\lambda$, which are used in triple correlations, is higher than that of the coefficients $a_\lambda$ used in the double cascades. Correlations that involve four or higher  observed transitions have not been explored, mainly because as the order of the type of correlation increases, the theoretical description becomes highly complex. 

The present method does not include attenuation/perturbation effects of the angular correlations due to relatively long half-lives (over 1 ns) of intermediate states~\cite{Hamilton_PhysRevC.5.899}. If, however, such an attenuation factor is measured for a particular cascade, it can be included in Eq.~\ref{eq: statistical tensor next simplified}, in order to correctly reproduce the attenuated angular correlation.

\subsection{Coupling with FIFRELIN code}

The present method can be coupled with powerful nuclear de-excitation codes such as FIFRELIN. FIFRELIN is a fission modelling code, which can simulate the de-excitation of every nucleus (not only a fission fragment) by neutron/gamma/conversion electron emission within the framework of nuclear realizations~\cite{BECVAR1998434}. The FIFRELIN code is used in various applications such the characterization of fission fragments~\cite{LITAIZE201251} as well in high energy physics simulations for the STEREO experiment~\cite{Allemandou_2018}. The full effect of directional $\gamma$ correlations can be a useful inclusion to simulations performed for such applications. 

As shown in Fig.~\ref{fig: 156Gd cascades}, the coupling of the present work with FIFRELIN is able to reproduce all angular correlations in the cascade. The present description can be applied to every isotope, both even and odd in mass number. For the moment, it is only restricted for $\gamma$-rays, but an implementation of other particle correlations, such as between internal conversion electrons, neutrons and gammas can be included in the future.


\subsection{Potential for use in the experimental determination of spins and mixing ratios}

Experimentally, the determination of the spin of a state is performed by measuring the experimental angular distribution $W_{exp}(\theta_{rel}^i)$ for a number of relative angles $\theta^i_{rel}$. These angles are usually defined by the experimental setup and the relative positions of the detectors used. Then, by correcting for the efficiency and the detectors finite size dimensions, a $\chi^2$-test is performed by forming the statistic:
\begin{equation}\label{eq:S_squared}
	S^2 = \sum_i \left[ \frac{W_{exp}(\theta_{rel}^i)-W_{th}(\delta,\theta^i_{rel}) }{\sigma_{W_i}} \right]^2
\end{equation}
for all possible spins and mixing ratios of the cascade. The values of spin and mixing ratios which best describe the experimental data are then adopted as the real values. There are other variations in the statistical treatment of this technique, and it is thoroughly discussed in~\cite{ROBINSON1990386}. It is important to note, that the correct application of this technique requires the determination of the correction factors $Q_\lambda$~\cite{KRANE1972_solid_angle_205,BARRETTE19711_solid_angle}, to account for the finite size dimensions of the detectors. These factors can be computed for simple co-axial geometries, but for complex geometries such those of clover detectors~\cite{Mutti_fipps_2018}, a simulation or additional measurements with known radioactive sources is necessary for their estimation.

Within the present approach, the simulation of the $\gamma$ cascades removes some important difficulties during the ``traditional" analysis of the data in such measurements. Hence, by simulating the angular correlations in a cascade of many $\gamma$-rays and using it as input for the simulation of the detector setup using GEANT4, a direct comparison can be then made between the simulation and the experimental data. In this case, there is no necessity to calculate or measure the geometrical correction factors for the detectors, as the simulation will directly produce the geometrical attenuation effect. The efficiency correction can also be excluded, unless there is significant deviation between the simulated efficiency and the measured one. In that case, the measured efficiency should be used as a parameter in the simulation. The $\chi^2$ minimization in Eq.~\ref{eq:S_squared} can be modified as:
\begin{equation}\label{eq:S_squared}
	S^2 = \sum_i \left[ \frac{W_{exp}(\theta_{rel}^i)-W_{sim}(\delta,\theta^i_{rel}) }{\sigma_{W_i}} \right]^2
\end{equation}
where now the simulated value $W_{sim}(\delta,\theta^i_{rel})$ has replaced the theoretical one.

The present approach also opens a new way for determining observables using triple or higher order angular correlations. As discussed in~\cite{Krane_1988_PhysRevC.37.747,KRANE1983321}, the sensitivity of the coefficients $\Gamma_k$ are more sensitive to the mixing ratio in some cases. Thus, triple correlations can sometimes offer an improved means of measurements. The direct comparison between simulation and measurements can make the analysis more time-efficient and avoids the use of complicated formulae, which are used for the triple angular correlations. Furthermore, the measurement of triple correlations is usually performed in standard detector geometries, where the theoretical description is simplified. This can pose difficulties in large detector arrays with fixed geometries. This restrictions are raised within the present framework, as a direct comparison between the simulated and experimental angular correlation can be made for triple angular correlations.

The present method is not restricted in angular correlations of a cascade of gammas which originates from an initial unoriented state. It can be also used in the case of oriented states produced for example, from nuclear reactions, provided that one knows the initial set of statistical tensors $\rho^\lambda_q(J_0)$. The initial condition of the statistical tensor must then be calculated in the framework of the reaction for each different case, or can be measured experimentally.

\section*{Acknowledgments}
We acknowledge the financial support of the Cross-Disciplinary Program on Numerical Simulation of CEA, the French Alternative Energies and Atomic Energy Commission.

\bibliography{ang_cor.bib}

\begin{thebibliography}{10}
\expandafter\ifx\csname url\endcsname\relax
  \def\url#1{\texttt{#1}}\fi
\expandafter\ifx\csname urlprefix\endcsname\relax\def\urlprefix{URL }\fi
\expandafter\ifx\csname href\endcsname\relax
  \def\href#1#2{#2} \def\path#1{#1}\fi

\bibitem{Hamilton1940_PhysRev.58.122}
D.~R. Hamilton, \href{https://link.aps.org/doi/10.1103/PhysRev.58.122}{On
  directional correlation of successive quanta}, Phys. Rev. 58 (1940) 122--131.
\newblock \href {http://dx.doi.org/10.1103/PhysRev.58.122}
  {\path{doi:10.1103/PhysRev.58.122}}.

\bibitem{Goertzel_1946_PhysRev.70.897}
G.~Goertzel, \href{https://link.aps.org/doi/10.1103/PhysRev.70.897}{Angular
  correlation of gamma-rays}, Phys. Rev. 70 (1946) 897--909.
\newblock \href {http://dx.doi.org/10.1103/PhysRev.70.897}
  {\path{doi:10.1103/PhysRev.70.897}}.

\bibitem{Brandy_PhysRev.78.558}
E.~L. Brady, M.~Deutsch,
  \href{https://link.aps.org/doi/10.1103/PhysRev.78.558}{Angular correlation of
  successive gamma-rays}, Phys. Rev. 78 (1950) 558--566.
\newblock \href {http://dx.doi.org/10.1103/PhysRev.78.558}
  {\path{doi:10.1103/PhysRev.78.558}}.

\bibitem{Biedenharn_Rose_RevModPhys.25.729}
L.~C. Biedenharn, M.~E. Rose,
  \href{https://link.aps.org/doi/10.1103/RevModPhys.25.729}{Theory of angular
  correlation of nuclear radiations}, Rev. Mod. Phys. 25 (1953) 729--777.
\newblock \href {http://dx.doi.org/10.1103/RevModPhys.25.729}
  {\path{doi:10.1103/RevModPhys.25.729}}.

\bibitem{Rose_Brink_Revmodphys_1967}
H.~J. Rose, D.~M. Brink, Angular distributions of gamma rays in terms of
  phase-defined reduced matrix elements, Rev. Mod. Phys. 39 (1967) 306--347.
\newblock \href {http://dx.doi.org/10.1103/RevModPhys.39.306}
  {\path{doi:10.1103/RevModPhys.39.306}}.

\bibitem{Fano_1957_RevModPhys.29.74}
U.~Fano, \href{https://link.aps.org/doi/10.1103/RevModPhys.29.74}{Description
  of states in quantum mechanics by density matrix and operator techniques},
  Rev. Mod. Phys. 29 (1957) 74--93.
\newblock \href {http://dx.doi.org/10.1103/RevModPhys.29.74}
  {\path{doi:10.1103/RevModPhys.29.74}}.

\bibitem{Hamilton1975_electromagnetic}
R.~Steffen, K.~Adler, W.~Hamilton~(Ed.), The electromagnetic interaction in
  nuclear spectroscopy, North-Holland, Amsterdam, 1975.

\bibitem{DEGROOT1952_1201}
S.~{De Groot},
  \href{https://www.sciencedirect.com/science/article/pii/S003189145280196X}{On
  the theories of angular distribution and correlation of beta and gamma
  radiation}, Physica 18~(12) (1952) 1201--1214.
\newblock \href
  {http://dx.doi.org/https://doi.org/10.1016/S0031-8914(52)80196-X}
  {\path{doi:https://doi.org/10.1016/S0031-8914(52)80196-X}}.

\bibitem{Kraus_1953}
A.~A. Kraus, Angular correlations of successive nuclear radiations,
  Dissertation (ph.d.), California Institute of Technology,
  doi:10.7907/KYA8-V204 (1953).

\bibitem{TAYLOR19711}
H.~Taylor, B.~Singh, F.~Prato, R.~McPherson,
  \href{https://www.sciencedirect.com/science/article/pii/S0092640X71800402}{A
  tabulation of gamma-gamma directional-correlation coefficients}, Atomic Data
  and Nuclear Data Tables 9~(1) (1971) 1--83.
\newblock \href
  {http://dx.doi.org/https://doi.org/10.1016/S0092-640X(71)80040-2}
  {\path{doi:https://doi.org/10.1016/S0092-640X(71)80040-2}}.

\bibitem{STUCHBERY200369}
A.~E. Stuchbery,
  \href{https://www.sciencedirect.com/science/article/pii/S0375947403011576}{γ-ray
  angular distributions and correlations after projectile-fragmentation
  reactions}, Nuclear Physics A 723~(1) (2003) 69--92.
\newblock \href
  {http://dx.doi.org/https://doi.org/10.1016/S0375-9474(03)01157-6}
  {\path{doi:https://doi.org/10.1016/S0375-9474(03)01157-6}}.

\bibitem{Litaize_2010_PhysRevC.82.054616}
O.~Litaize, O.~Serot,
  \href{https://link.aps.org/doi/10.1103/PhysRevC.82.054616}{Investigation of
  phenomenological models for the monte carlo simulation of the prompt fission
  neutron and $\ensuremath{\gamma}$ emission}, Phys. Rev. C 82 (2010) 054616.
\newblock \href {http://dx.doi.org/10.1103/PhysRevC.82.054616}
  {\path{doi:10.1103/PhysRevC.82.054616}}.

\bibitem{LITAIZE201251}
O.~Litaize, O.~Serot, D.~Regnier, S.~Theveny, S.~Onde,
  \href{https://www.sciencedirect.com/science/article/pii/S1875389212011947}{New
  features of the fifrelin code for the investigation of fission fragments
  characteristics}, Physics Procedia 31 (2012) 51--58, gAMMA-1 Emission of
  Prompt Gamma-Rays in Fission and Related Topics.
\newblock \href {http://dx.doi.org/https://doi.org/10.1016/j.phpro.2012.04.008}
  {\path{doi:https://doi.org/10.1016/j.phpro.2012.04.008}}.

\bibitem{Litaize2015}
O.~Litaize, O.~Serot, L.~Berge,
  \href{https://www.sciencedirect.com/science/article/pii/0010465583901182}{Fission
  modelling with fifrelin}, The European Physics Journal A 51 (2015) 117.
\newblock \href {http://dx.doi.org/https://doi.org/10.1140/epja/i2015-15177-9}
  {\path{doi:https://doi.org/10.1140/epja/i2015-15177-9}}.

\bibitem{REGNIER201619}
D.~Regnier, O.~Litaize, O.~Serot,
  \href{https://www.sciencedirect.com/science/article/pii/S0010465515004452}{An
  improved numerical method to compute neutron/gamma deexcitation cascades
  starting from a high spin state}, Computer Physics Communications 201 (2016)
  19--28.
\newblock \href {http://dx.doi.org/https://doi.org/10.1016/j.cpc.2015.12.007}
  {\path{doi:https://doi.org/10.1016/j.cpc.2015.12.007}}.

\bibitem{BECVAR1998434}
F.~Bečvář,
  \href{https://www.sciencedirect.com/science/article/pii/S0168900298007876}{Simulation
  of γ cascades in complex nuclei with emphasis on assessment of uncertainties
  of cascade-related quantities}, Nuclear Instruments and Methods in Physics
  Research Section A: Accelerators, Spectrometers, Detectors and Associated
  Equipment 417~(2) (1998) 434--449.
\newblock \href
  {http://dx.doi.org/https://doi.org/10.1016/S0168-9002(98)00787-6}
  {\path{doi:https://doi.org/10.1016/S0168-9002(98)00787-6}}.

\bibitem{CAPOTE20093107}
R.~Capote, M.~Herman, P.~Obložinský, P.~Young, S.~Goriely, T.~Belgya,
  A.~Ignatyuk, A.~Koning, S.~Hilaire, V.~Plujko, M.~Avrigeanu, O.~Bersillon,
  M.~Chadwick, T.~Fukahori, Z.~Ge, Y.~Han, S.~Kailas, J.~Kopecky, V.~Maslov,
  G.~Reffo, M.~Sin, E.~Soukhovitskii, P.~Talou,
  \href{https://www.sciencedirect.com/science/article/pii/S0090375209000994}{Ripl
  – reference input parameter library for calculation of nuclear reactions
  and nuclear data evaluations}, Nuclear Data Sheets 110~(12) (2009)
  3107--3214, special Issue on Nuclear Reaction Data.
\newblock \href {http://dx.doi.org/https://doi.org/10.1016/j.nds.2009.10.004}
  {\path{doi:https://doi.org/10.1016/j.nds.2009.10.004}}.

\bibitem{Allemandou_2018}
N.~Allemandou, H.~Almaz{\'{a}}n, P.~del Amo~Sanchez, L.~Bernard, C.~Bernard,
  A.~Blanchet, A.~Bonhomme, G.~Bosson, O.~Bourrion, J.~Bouvier, C.~Buck,
  V.~Caillot, M.~Chala, P.~Champion, P.~Charon, A.~Collin, P.~Contrepois,
  G.~Coulloux, B.~Desbri{\`{e}}res, G.~Deleglise, W.~E. Kanawati, J.~Favier,
  S.~Fuard, I.~G. Monteiro, B.~Gramlich, J.~Haser, V.~Helaine, M.~Heusch,
  M.~Jentschel, F.~Kandzia, G.~Konrad, U.~Köster, S.~Kox, C.~Lahonde-Hamdoun,
  J.~Lamblin, A.~Letourneau, D.~Lhuillier, C.~Li, M.~Lindner, L.~Manzanillas,
  T.~Materna, O.~M{\'{e}}plan, A.~Minotti, C.~Monon, F.~Montanet, F.~Nunio,
  F.~Peltier, Y.~Penichot, M.~Pequignot, H.~Pessard, Y.~Piret, G.~Prono,
  G.~Qu{\'{e}}m{\'{e}}ner, J.-S. Real, C.~Roca, T.~Salagnac, V.~Sergeyeva,
  S.~Schoppmann, L.~Scola, J.-P. Scordilis, T.~Soldner, A.~Stutz, D.~Tourres,
  C.~Vescovi, S.~Zsoldos,
  \href{https://doi.org/10.1088/1748-0221/13/07/p07009}{The {STEREO}
  experiment}, Journal of Instrumentation 13~(07) (2018) P07009--P07009.
\newblock \href {http://dx.doi.org/10.1088/1748-0221/13/07/p07009}
  {\path{doi:10.1088/1748-0221/13/07/p07009}}.

\bibitem{Almazan2019}
H.~Almaz{\'a}n, L.~Bernard, A.~Blanchet, A.~Bonhomme, C.~Buck, A.~Chebboubi,
  P.~del Amo~Sanchez, I.~El~Atmani, J.~Haser, F.~Kandzia, S.~Kox, L.~Labit,
  J.~Lamblin, A.~Letourneau, D.~Lhuillier, M.~Lindner, O.~Litaize, T.~Materna,
  A.~Minotti, H.~Pessard, J.-S. R{\'e}al, C.~Roca, T.~Salagnac, V.~Savu,
  S.~Schoppmann, V.~Sergeyeva, T.~Soldner, A.~Stutz, L.~Thulliez, M.~Vialat,
  \href{https://doi.org/10.1140/epja/i2019-12886-y}{Improved stereo simulation
  with a new gamma ray spectrum of excited gadolinium isotopes using fifrelin},
  The European Physical Journal A 55~(10) (2019) 183.
\newblock \href {http://dx.doi.org/10.1140/epja/i2019-12886-y}
  {\path{doi:10.1140/epja/i2019-12886-y}}.

\bibitem{Geant4_ALLISON2016186}
J.~Allison, K.~Amako, J.~Apostolakis, P.~Arce, M.~Asai, T.~Aso, E.~Bagli,
  A.~Bagulya, S.~Banerjee, G.~Barrand, B.~Beck, A.~Bogdanov, D.~Brandt,
  J.~Brown, H.~Burkhardt, P.~Canal, D.~Cano-Ott, S.~Chauvie, K.~Cho,
  G.~Cirrone, G.~Cooperman, M.~Cortés-Giraldo, G.~Cosmo, G.~Cuttone,
  G.~Depaola, L.~Desorgher, X.~Dong, A.~Dotti, V.~Elvira, G.~Folger,
  Z.~Francis, A.~Galoyan, L.~Garnier, M.~Gayer, K.~Genser, V.~Grichine,
  S.~Guatelli, P.~Guèye, P.~Gumplinger, A.~Howard, I.~Hřivnáčová,
  S.~Hwang, S.~Incerti, A.~Ivanchenko, V.~Ivanchenko, F.~Jones, S.~Jun,
  P.~Kaitaniemi, N.~Karakatsanis, M.~Karamitros, M.~Kelsey, A.~Kimura, T.~Koi,
  H.~Kurashige, A.~Lechner, S.~Lee, F.~Longo, M.~Maire, D.~Mancusi, A.~Mantero,
  E.~Mendoza, B.~Morgan, K.~Murakami, T.~Nikitina, L.~Pandola, P.~Paprocki,
  J.~Perl, I.~Petrović, M.~Pia, W.~Pokorski, J.~Quesada, M.~Raine, M.~Reis,
  A.~Ribon, A.~{Ristić Fira}, F.~Romano, G.~Russo, G.~Santin, T.~Sasaki,
  D.~Sawkey, J.~Shin, I.~Strakovsky, A.~Taborda, S.~Tanaka, B.~Tomé,
  T.~Toshito, H.~Tran, P.~Truscott, L.~Urban, V.~Uzhinsky, J.~Verbeke,
  M.~Verderi, B.~Wendt, H.~Wenzel, D.~Wright, D.~Wright, T.~Yamashita,
  J.~Yarba, H.~Yoshida,
  \href{https://www.sciencedirect.com/science/article/pii/S0168900216306957}{Recent
  developments in geant4}, Nuclear Instruments and Methods in Physics Research
  Section A: Accelerators, Spectrometers, Detectors and Associated Equipment
  835 (2016) 186--225.
\newblock \href {http://dx.doi.org/https://doi.org/10.1016/j.nima.2016.06.125}
  {\path{doi:https://doi.org/10.1016/j.nima.2016.06.125}}.

\bibitem{Singh_PhysRevC.4.1510}
B.~P. Singh, H.~S. Dahiya, U.~S. Pande,
  \href{https://link.aps.org/doi/10.1103/PhysRevC.4.1510}{Sign of
  $\ensuremath{\delta}$, the amplitude mixing ratio of gamma transitions, and
  beta-gamma-gamma and gamma-gamma-gamma angular-correlation studies}, Phys.
  Rev. C 4 (1971) 1510--1513.
\newblock \href {http://dx.doi.org/10.1103/PhysRevC.4.1510}
  {\path{doi:10.1103/PhysRevC.4.1510}}.

\bibitem{Krane_1988_PhysRevC.37.747}
K.~S. Krane, N.~S. Schulz,
  \href{https://link.aps.org/doi/10.1103/PhysRevC.37.747}{Triple angular
  correlations in the decay of $^{110}$${\mathrm{ag}}^{\mathrm{m}}$}, Phys.
  Rev. C 37 (1988) 747--753.
\newblock \href {http://dx.doi.org/10.1103/PhysRevC.37.747}
  {\path{doi:10.1103/PhysRevC.37.747}}.

\bibitem{KRANE1983321}
K.~Krane,
  \href{https://www.sciencedirect.com/science/article/pii/0167508783905999}{Some
  remarks concerning triple angular correlations}, Nuclear Instruments and
  Methods in Physics Research 214~(2) (1983) 321--332.
\newblock \href
  {http://dx.doi.org/https://doi.org/10.1016/0167-5087(83)90599-9}
  {\path{doi:https://doi.org/10.1016/0167-5087(83)90599-9}}.

\bibitem{KRANE_steffen_wheeler_1973351}
K.~Krane, R.~Steffen, R.~Wheeler,
  \href{https://www.sciencedirect.com/science/article/pii/S0092640X73800166}{Directional
  correlations of gamma radiations emitted from nuclear states oriented by
  nuclear reactions or cryogenic methods}, Atomic Data and Nuclear Data Tables
  11~(5) (1973) 351--406.
\newblock \href
  {http://dx.doi.org/https://doi.org/10.1016/S0092-640X(73)80016-6}
  {\path{doi:https://doi.org/10.1016/S0092-640X(73)80016-6}}.

\bibitem{KRANE1972_solid_angle_205}
K.~Krane,
  \href{https://www.sciencedirect.com/science/article/pii/0029554X72900997}{Solid-angle
  correction factors for coaxial ge(li) detectors}, Nuclear Instruments and
  Methods 98~(2) (1972) 205--210.
\newblock \href
  {http://dx.doi.org/https://doi.org/10.1016/0029-554X(72)90099-7}
  {\path{doi:https://doi.org/10.1016/0029-554X(72)90099-7}}.

\bibitem{Krane1973SolidangleCF}
K.~Krane, Solid-angle correction factors for five-sided coaxial ge(li)
  detectors, Nuclear Instruments and Methods 109 (1973) 401--402.

\bibitem{CAMP1969192}
D.~Camp, A.~{Van Lehn},
  \href{https://www.sciencedirect.com/science/article/pii/0029554X69900238}{Finite
  solid-angle corrections for ge(li) detectors}, Nuclear Instruments and
  Methods 76~(2) (1969) 192--240.
\newblock \href
  {http://dx.doi.org/https://doi.org/10.1016/0029-554X(69)90023-8}
  {\path{doi:https://doi.org/10.1016/0029-554X(69)90023-8}}.

\bibitem{Moura_2019_doi:10.1119/1.5099891}
F.~Moura, \href{https://doi.org/10.1119/1.5099891}{Maximal angular correlation
  in γ – γ coincidences: A quantitative study}, American Journal of Physics
  87~(8) (2019) 638--642.
\newblock \href {http://arxiv.org/abs/https://doi.org/10.1119/1.5099891}
  {\path{arXiv:https://doi.org/10.1119/1.5099891}}, \href
  {http://dx.doi.org/10.1119/1.5099891} {\path{doi:10.1119/1.5099891}}.

\bibitem{AMSLER198321}
C.~Amsler, J.~Bizot,
  \href{https://www.sciencedirect.com/science/article/pii/0010465583901182}{Simulation
  of angular distributions and correlations in the decay of particles with
  spin}, Computer Physics Communications 30~(1) (1983) 21--30.
\newblock \href
  {http://dx.doi.org/https://doi.org/10.1016/0010-4655(83)90118-2}
  {\path{doi:https://doi.org/10.1016/0010-4655(83)90118-2}}.

\bibitem{Ponkratenko2000}
O.~A. Ponkratenko, V.~I. Tretyak, Y.~G. Zdesenko,
  \href{https://doi.org/10.1134/1.855784}{Event generator decay4 for simulating
  double-beta processes and decays of radioactive nuclei}, Physics of Atomic
  Nuclei 63~(7) (2000) 1282--1287.
\newblock \href {http://dx.doi.org/10.1134/1.855784}
  {\path{doi:10.1134/1.855784}}.

\bibitem{geant_4_sim_2017}
M.~Buuck, J.~Detwiler, I.~Guinn, A.~Li,
  \href{https://indico.cern.ch/event/627270/contributions/2534683/attachments/1441154/2218794/170406_Guinn_RDM_workshop.pdf}{Angular
  correlations in gamma deexcitation}, geant 4 RDM Mini Workshop (2017).
\newline\urlprefix\url{https://indico.cern.ch/event/627270/contributions/2534683/attachments/1441154/2218794/170406_Guinn_RDM_workshop.pdf}

\bibitem{Turner_2020}
A.~Turner, C.~Wheldon, M.~Gilbert, L.~Packer, J.~Burns, T.~K. Wheldon,
  M.~Freer, \href{https://doi.org/10.1088/1742-6596/1643/1/012211}{Generalised
  gamma spectrometry simulator for problems in nuclide identification}, Journal
  of Physics: Conference Series 1643 (2020) 012211.
\newblock \href {http://dx.doi.org/10.1088/1742-6596/1643/1/012211}
  {\path{doi:10.1088/1742-6596/1643/1/012211}}.

\bibitem{SMITH201947}
J.~Smith, A.~MacLean, W.~Ashfield, A.~Chester, A.~Garnsworthy, C.~Svensson,
  \href{https://www.sciencedirect.com/science/article/pii/S0168900218314116}{Gamma–gamma
  angular correlation analysis techniques with the griffin spectrometer},
  Nuclear Instruments and Methods in Physics Research Section A: Accelerators,
  Spectrometers, Detectors and Associated Equipment 922 (2019) 47--63.
\newblock \href {http://dx.doi.org/https://doi.org/10.1016/j.nima.2018.10.097}
  {\path{doi:https://doi.org/10.1016/j.nima.2018.10.097}}.

\bibitem{Snelling_1983}
D.~M. Snelling, W.~D. Hamilton,
  \href{https://doi.org/10.1088/0305-4616/9/1/016}{Gamma-gamma directional
  correlation measurements in146nd following thermal-neutron capture}, Journal
  of Physics G: Nuclear Physics 9~(1) (1983) 111--130.
\newblock \href {http://dx.doi.org/10.1088/0305-4616/9/1/016}
  {\path{doi:10.1088/0305-4616/9/1/016}}.

\bibitem{abramovitz_handbook_1964}
M.~Abramovitz, I.~Stegun, Handbook of Mathematical Functions, National Bureau
  of Standards, Applied Mathematics Series 55, 1964, Ch. 27.9.

\bibitem{Wigner1993}
E.~P. Wigner, \href{https://doi.org/10.1007/978-3-662-02781-3_42}{On the
  Matrices Which Reduce the Kronecker Products of Representations of S. R.
  Groups}, Springer Berlin Heidelberg, Berlin, Heidelberg, 1993, pp. 608--654.
\newblock \href {http://dx.doi.org/10.1007/978-3-662-02781-3_42}
  {\path{doi:10.1007/978-3-662-02781-3_42}}.
\newline\urlprefix\url{https://doi.org/10.1007/978-3-662-02781-3_42}

\bibitem{KRANE1975_mixing_ratio_383}
K.~S. Krane,
  \href{https://www.sciencedirect.com/science/article/pii/0092640X75900182}{E2,m1
  multipole mixing ratios in even-even nuclei, a ≥ 152}, Atomic Data and
  Nuclear Data Tables 16~(4) (1975) 383--408.
\newblock \href
  {http://dx.doi.org/https://doi.org/10.1016/0092-640X(75)90018-2}
  {\path{doi:https://doi.org/10.1016/0092-640X(75)90018-2}}.

\bibitem{Hamilton_1985}
J.~H. Hamilton, P.~G. Hansen, E.~F. Zganjar, Advances in studies of nuclei far
  from stability, Reports on Progress in Physics 48~(5) (1985) 631--708.
\newblock \href {http://dx.doi.org/10.1088/0034-4885/48/5/002}
  {\path{doi:10.1088/0034-4885/48/5/002}}.

\bibitem{Ferentz_F_coeff}
M.~Ferentz, N.~Rosenzweig, Anl-5324, Available from Clearinghouse for Federal
  Scientific and Technical Information, U.S. Dept. of Commerce, Springfield,
  Va. 22151.

\bibitem{Blin-Stoyle1957}
R.~J. Blin-Stoyle, M.~A. Grace,
  \href{https://doi.org/10.1007/978-3-642-45878-1_6}{Oriented Nuclei}, Springer
  Berlin Heidelberg, Berlin, Heidelberg, 1957, pp. 555--610.
\newblock \href {http://dx.doi.org/10.1007/978-3-642-45878-1_6}
  {\path{doi:10.1007/978-3-642-45878-1_6}}.
\newline\urlprefix\url{https://doi.org/10.1007/978-3-642-45878-1_6}

\bibitem{ang_corr_calc}
"GRIFFIN-collaboration", {Angular Correlation Utility},
  \url{http://griffincollaboration.github.io/AngularCorrelationUtility/},
  [Online; accessed 2020] (2016).
\newblock \href {http://dx.doi.org/10.5281/zenodo.45587}
  {\path{doi:10.5281/zenodo.45587}}.

\bibitem{Hamilton_PhysRevC.5.899}
J.~H. Hamilton, P.~E. Little, A.~V. Ramayya, E.~Collins, N.~R. Johnson, J.~J.
  Pinajian, A.~F. Kluk,
  \href{https://link.aps.org/doi/10.1103/PhysRevC.5.899}{$m1$ admixtures of
  transitions in $^{156}\mathrm{Gd}$}, Phys. Rev. C 5 (1972) 899--910.
\newblock \href {http://dx.doi.org/10.1103/PhysRevC.5.899}
  {\path{doi:10.1103/PhysRevC.5.899}}.

\bibitem{ROBINSON1990386}
S.~Robinson, How reliable are spins and δ-values derived from directional
  correlation experiments?, Nuclear Instruments and Methods in Physics Research
  Section A: Accelerators, Spectrometers, Detectors and Associated Equipment
  292~(2) (1990) 386 -- 400.
\newblock \href
  {http://dx.doi.org/https://doi.org/10.1016/0168-9002(90)90395-M}
  {\path{doi:https://doi.org/10.1016/0168-9002(90)90395-M}}.

\bibitem{BARRETTE19711_solid_angle}
J.~Barrette, G.~Lamoureux, S.~Monaro,
  \href{https://www.sciencedirect.com/science/article/pii/0029554X71901315}{Geometrical
  correction factors for angular correlation measurements with ge(li)
  detectors}, Nuclear Instruments and Methods 93~(1) (1971) 1--11.
\newblock \href
  {http://dx.doi.org/https://doi.org/10.1016/0029-554X(71)90131-5}
  {\path{doi:https://doi.org/10.1016/0029-554X(71)90131-5}}.

\bibitem{Mutti_fipps_2018}
{Mutti, P.}, {Blanc, A.}, {Jentschel, M.}, {K\"oster, U.}, {Michelagnoli, C.},
  {Ruiz-Martinez, E.},
  \href{https://doi.org/10.1051/epjconf/201817001013}{Fipps: Fission product
  prompt gamma-ray spectrometer, a new high efficiency hpge array for nuclear
  spectroscopy}, EPJ Web Conf. 170 (2018) 01013.
\newblock \href {http://dx.doi.org/10.1051/epjconf/201817001013}
  {\path{doi:10.1051/epjconf/201817001013}}.

\end{thebibliography}
\end{document}